\documentclass[12pt]{article}
\usepackage{epsfig}
\usepackage{tabularx}
\usepackage[titletoc, title]{appendix}
\usepackage{setspace}
\usepackage{lscape}
\usepackage{booktabs}
\usepackage{array}
\usepackage[english]{babel}
\usepackage{makeidx}
\usepackage{latexsym}
\usepackage{amssymb}
\usepackage{qsymbols}
\usepackage{graphicx}

\usepackage{amsmath, amsthm, amssymb}
\usepackage{nomencl}
\usepackage{float}
\usepackage{multirow}
\usepackage{multicol}
\usepackage{xcolor,soul}
\usepackage{lscape}
\usepackage{tabularx}
\usepackage{pdflscape}
\usepackage{epstopdf}
\usepackage{color}
\usepackage{dsfont}
\usepackage{natbib}

\providecommand{\keywords}[1]{\textbf{\textit{Keywords:}} #1}

\usepackage{changes}
\definechangesauthor[name={Marco}, color=red]{mal}
\definechangesauthor[name={Nicola}, color=blue]{nic}
\definechangesauthor[name={Giovanna}, color=violet]{gio}

\newcommand{\beq}{\begin{equation}}
\newcommand{\eeq}{\end{equation}}
\newcommand{\bed}{\begin{displaymath}}
\newcommand{\eed}{\end{displaymath}}

\newcommand{\bs}{\boldsymbol } 
\newcommand{\tbf}{\textbf } 
   
\def\b#1{\mbox{\boldmath $#1$}}

\newcommand{\bbeta}{\boldsymbol{\beta}}
\newcommand{\blambda}{\boldsymbol{\lambda}}

\def \bPhi {\boldsymbol\Phi}

\title{M-quantile regression for {multivariate} longitudinal data: analysis of the Millennium Cohort Study data}

\author{Marco Alf\`o\thanks{Dipartimento di Scienze Statistiche, Sapienza Universit\`a di Roma \texttt{marco.alfo@uniroma1.it}} \and Maria Francesca Marino\thanks{Dipartimento di Statistica, Informatica, Applicazioni, Universit\`a  degli Studi di Firenze} \and Maria Giovanna Ranalli\thanks{Dipartimento di Scienze Politiche, Universit\`a  degli Studi di Perugia} \and Nicola Salvati\thanks{Dipartimento di Economia e Management, Universit\`a di Pisa} \and Nikos Tzavidis\thanks{Department of Social Statistics and Demography, Southampton Statistical Sciences Research Institute, University of Southampton}}

\begin{document}

\maketitle

\begin{abstract}
We propose a M-quantile regression model for the analysis of multivariate, continuous, longitudinal data. M-quantile regression represents an appealing alternative to standard regression models, as it combines the robustness of quantile and the efficiency of expectile regression, providing a complete picture  of the response variable distribution. 
Discrete, individual-specific, random parameters are used to account for both dependence within the same response recorded at different times and association between different responses observed on the same sample unit at a given time. A suitable parametrisation is also introduced in the linear predictor to account for possible dependence between the individual specific random parameters and the vector of observed covariates, that is to account for \emph{endogeneity} of some covariates. An extended EM algorithm is proposed to derive model parameter estimates under a maximum likelihood approach. {The model is applied to the analysis of the strengths and difficulties questionnaire scores from the Millennium Cohort Study in the UK}. 
\bigskip

\end{abstract}


\section{Introduction}\label{sec_intro}
The analysis of longitudinal data has represented in the last years an interesting and rapidly increasing field of research as it allows to obtain in-depth information about the evolution of phenomena over time. When dealing with such studies, some form of dependence between observations from the same individual must be taken into account to avoid misleading inferential conclusions. In a regression framework, sources of unobserved heterogeneity are typically used to account for such a feature. In particular, individual-specific random parameters may be considered as part of the model specification to account for a (basic) form of dependence. The resulting model is known, in the literature, as random or mixed effect model; for early developments see \cite{LairdWare1982}.

Recently, there has been an increasing interest in the analysis of longitudinal data via quantile regression models. In this framework, the aim is to study the effect of observed covariates on different regions of the response variable distribution. As a result, when compared to mean regression, a more a complete picture of this distribution may be obtained \citep{Kneib2013}.  To deal with longitudinal observations, \cite{GeraciBottai2007} introduced a quantile regression model with individual-specific random intercepts distributed according to a Gaussian or an Asymmetric Laplace density. \citet{LiuBottai2009} and \citet{GeraciBottai2013} further extended the model to deal with general random parameters. \cite{AlfoRanalliSalvati2016} considered a quantile regression model with individual-specific random parameters having unspecified distribution directly estimated from the observed data. 
\cite{Farcomeni2012} introduced hidden Markov quantile regression with individual-specific random intercepts evolving over time according to a hidden Markov chain. \cite{Marino2016} considered a mixed hidden Markov quantile regression model to account for both time-constant and time-varying random parameters. For a recent review of available approaches to quantile regression for longitudinal data, see \cite{MarinoFarco2015}.

A further generalization of quantile regression is provided by M-quantile regression \citep{Brec:Cham:1988}. In this framework, the conditional distribution of the response variable is characterized in terms of different location parameters, the M-quantiles. Although these have a less intuitive interpretation than standard quantiles \citep{Jon94}, M-quantile regression offers a number of advantages: (i) it easily allows for robust estimation; (ii) it can trade robustness and efficiency in inference by selecting the tuning constant of the influence function; (iii) it offers computational stability due to the wide range of available continuous influence functions with respect to the more standard absolute value used in the quantile regression context \citep{Tzavidis2016}.  

The extension of M-quantile regression to a longitudinal data context is quite recent. \cite{Tzavidis2016} proposed a model based on individual-specific random intercepts to account for the dependence between measures coming from the same unit. Using a semi-parametric approach, \cite{AlfoRanalliSalvati2016} considered a finite mixture of M-quantile regression models.
In this paper, we extend this latter approach to the analysis of multivariate longitudinal responses. It is worth to notice that this extension is different from the multivariate M-quantile approach proposed by \cite{Brec:2001} which, instead, is a probability-based order technique for summarizing the distribution of multivariate data. When compared to univariate analysis, multivariate regression models may lead to more efficient estimates and may help obtain a measure of correlation between the responses. 

A standard assumption of mixed effect models is that individual-specific random parameters represent sources of unobserved heterogeneity that summarize the effect of \emph{omitted} covariates on the response distribution. While it is well known that this assumption may lead to a flexible modelling specification, possible consequences are not always properly handled. In fact, unobserved heterogeneity may be correlated with the observed covariates, leading to a form of \emph{endogeneity} which may cause the corresponding ML estimators to be inconsistent. To handle dependence between random parameters and observed covariates, we propose an \emph{auxiliary} regression model by extending the proposal by \cite{Mundlak1978} and \cite{Chamberlain1980, Chamberlain1984} to the context of M-quantile regression. Related approaches in the quantile regression framework are those by \cite{smithetal2015}, \cite{arellanobonhomme2016} and  \cite{weidnermoon2016}.
To the best of our knowledge, the present manuscript represents the first attempt to account for endogeneity in the M-quantile literature as well as the first to deal with multivariate responses. We 

Our proposal is applied to data on emotional and behavioral problems of children participating to the Millennium Cohort Study (MCS), a multi-disciplinary research project following the lives of around $19,$000 children born in the UK in $2000/01$. 
\cite{Tzavidis2016} analysed the effect of neighbourhood and family risk factors on these two outcomes
via univariate M-quantile regression models, considering individual-specific random effects to account for dependence between repeated measures from the same unit and pseudo-BLUP equations \citep[see e.g.][]{Harville1976} to derive parameter estimates. Here, we analyse the same dataset using a finite mixture of M-quantile regression models for multivariate responses, handling possible endogeneity of observed covariates via an \emph{auxiliary} regression model. 
 
The paper is organized as follows: in Section \ref{MCS_study}, we describe the Millennium Cohort Study data. In Section \ref{sec1.mixedQR}, we extend the M-quantile finite mixture approach by \cite{AlfoRanalliSalvati2016} to multivariate longitudinal data. Section \ref{endog} presents a possible solution to potential endogeneity in M-quantile regression models. Section \ref{semip} shows the proposed semi-parametric approach to M-quantile models for multivariate longitudinal responses. In Section \ref{sec:4}, we discuss parameter estimation and inference. Section \ref{sec:7} presents the results from the analysis of the Millennium Cohort Study data. The last section contains concluding remarks and outlines a potential future research agenda.

\section{The Millennium Cohort Study data}\label{MCS_study}
The Millennium Cohort Study is a longitudinal study conducted in the United Kingdom. It involves children who were born between September 1, 2000 and August 31, 2001 in England and Wales, or between November 24, 2000 and January 11, 2002 in Scotland and Northern Ireland, living in the UK at nine months of age, whose families were eligible to receive Child Benefit at that age. The study was designed to over-represent children from deprived backgrounds, with the aim at better addressing the effects of social disadvantage on children's outcomes. For this purpose, and in order to account for the increasing diversity in the UK population with respect to differential health, education and social access, the study was also designed to over-represent areas with high ethnic minority concentration. For England and Wales, the population was stratified into three different strata: the first stratum, \textit{ethnic minority}, comprises children living in wards were the proportion of ethnic minorities was not less than $30 \%$ in the $1991$ Census; the second stratum, \textit{disadvantaged}, comprises children living in wards, other than those falling into the previous stratum, which fell into the poorest $25\%$ wards based on the Child Poverty Index; the latter stratum, \textit{advantaged}, comprises children not included in the previous two. For Wales, Scotland, and Northern Ireland, due to the low percentages of ethnic minority groups, the population was stratified into two groups only: the \textit{disadvantaged} and the \textit{advantaged} stratum.

The MCS sample was randomly selected within each stratum and each country. Once the sample wards were selected, a list of all children turning nine months old during the survey window and living in the selected wards was generated. Overall, a cohort of $18,$818 children was eligible. These children were followed for up to five time periods. The first measurement took place when children were about $9$ months old; subsequent measures were recorder at $3, 5, 7$, and $11$ years of age. 

Children's emotional and behavioral problems were measured by means of the so called internalizing SDQ (i-SDQ) and externalizing SDQ scores (e-SDQ), respectively. These were obtained by the sum of responses given by the main children's caregiver to a series of items that describe children's problems. The $25$-item SDQ comprises five domains measured by five items each: emotional symptoms, peer problems, conduct problems, hyperactivity, and pro-social behaviour. For each item, a $0$ score is given if the response is not true, $1$ if it is somewhat true and $2$ if it is certainly true. The internalizing SDQ (i-SDQ) score is the sum of responses to the five emotional symptom items and the five peer problems items. Therefore, it ranges in the interval $\left[0,20\right]$. The externalizing SDQ (e-SDQ) score is the sum of responses to the five conduct problem items and the five hyperactivity items (also in this case the range is $\left[0,20\right]$). 

Internalizing and externalizing scores were recorded when children were $3,5,$ and $7$ years old. 
In this paper, we use data on $9,021$ children who were present to at least one of these measurement occasions. 
In particular, data from $7,$055 children are available for the first wave, data from $7,$938 children are available for wave $2$, and data from $7,$078 children are available at the third one. That is, 
$5,$342 units ($59.22\%$) present complete data records, while the remaining ones ($40.78\%$) have incomplete information. As it is frequently done when dealing with longitudinal observations subject to missingness, we assume a MAR mechanism to generate the observed responses.
To analyse the effect of demographic and socio-economic factors on the two outcomes above, following \cite{Tzavidis2016}, we decided to focus on the covariates described in the following.
\begin{itemize}
\item $\mbox{ALE}_{11}$: number of potentially Adverse (stressful) Life Events experienced by the family between two consecutive waves. This variable was obtained as the sum of responses to the $11$ items of the adverse life event scale by \cite{Tiet1998}. 
That is: a family member died, a negative change in the financial situation was experienced, a new step-parent came, a sibling left home, the child got seriously sick or injured, a divorce or a separation took place, the family moved, a parent lost the job, a new natural sibling came, a new step-sibling came, the mother was diagnosed with or treated for depression.

\item $\mbox{SED}_{4}$: measure of Socio-Economic Disadvantage, obtained as the sum of responses to four items on family poverty. Such items were defined as:  overcrowding (more than 1.5 people per room excluding bathroom and kitchen), not owning a home, receipt of an income support, annual income below the poverty line defined as the $60\%$ of the UK national median household income.

\item Kessm: measure of maternal depression based on the Kessler scale \citep{Kessler1992}, with a range in $\left[0, 24\right]$; higher values identify more severe depression symptoms. 

\item IMD: measure of neighbourhood deprivation based on the Index of Multiple Deprivation,  with a range in $\left[1, 10 \right]$; lower values correspond to areas with higher deprivation.
\end{itemize}

\noindent Within the set of covariates, we also included child age (Age), maternal education (no qualification, degree, GCSE), ethnicity (non-white and white), gender, and the stratification variable.

As pointed out above, the scientific question entails the effect of neighbourhood and family risk factors on children emotional and behavioral problems, as measured by the i-SDQ and the e-SDQ scores. We should remark two issues. First, it is possible that the effect of some risk factors on the SDQ scores is not  constant across the corresponding distribution. Second, in this particular application, our interest mainly relies on understanding the effect of observed covariates on the right tail of the distribution, which is associated with more problematic children. As we are dealing with longitudinal data, dependence between measures from the same child needs to be taken into account to obtain valid inference. In the following, we will present a model specification which properly answers to all of these issues. 

\section{M-quantile regression for multivariate longitudinal responses} \label{sec1.mixedQR}

Let $Y_{ith}$ denote {a continuous longitudinal random variable and $y_{ith}$ the corresponding observed value for the $i$-th individual, $i=1, \dots, n$, at time occasions $t = 1,\dots,T_i,$ for the the $h$-th outcome, $h = 1, \dots, H$. Furthermore, let $\b x_{ith}$ represent a $p$-dimensional vector of observed covariates. We are interested in analysing how these covariates influence the distribution of the observed responses. 
As it is frequent in the longitudinal data literature, association between observations from the same individual is modelled via individual-specific random parameters that describe potential sources of unobserved heterogeneity between individuals under observation. 
For the $h$-th outcome $Y_{ith}$ and for a given M-quantile $q \in (0,1)$, we assume that, conditional on a $s$-dimensional vector of random parameters ${\bf b}_{ih,q}$ and the {$(T_i\times p )$}-dimensional matrix of covariates $\tbf X_{ih}$, measurements from the same unit are independent (\emph{local} or \emph{conditional} independence).
Based on such an assumption, the \emph{conditional} density for the individual sequence referring to a given outcome $h= 1, \dots, H,$ could be written as follows:
\beq\label{uniconddens}
f_{q}\left({\bf y}_{ih} \mid {\bf b}_{ih,q}, {\bf X}_{ih}\right)= \prod_{t=1}^{T_{i}}f_{q}\left(y_{ith} \mid {\bf b}_{ih,q}, {\bf x}_{ith}\right),
\eeq
where ${\bf y}_{ih}$ denotes the $T_i$-dimensional vector of observed responses for the $i$-th individual and the $h$-th response variable.
When we move from considering repeated measures for a single outcome as in (\ref{uniconddens}) to considering repeated measures for a multivariate outcome, that is random variables  $\b Y_{i} = ({\bf Y}_{i1}, \dots, {\bf Y}_{iH})$, we should take into account that the corresponding elements may be dependent. To describe such a dependence, we introduce a further local independence assumption. We assume that, \emph{conditional} on the outcome-specific random parameters ${\bf b}_{ih,q}$ and the matrix of individual covariates ${\bf X}_{ih}$, the conditional density for the multivariate response associated to the $i$-th statistical unit may be written as follows:
\beq\label{mulconddens}
f_{q}\left({\bf y}_{i} \mid {\bf b}_{i,q}, {\bf X}_{i}\right)= \prod_{h=1}^{H}f_{q}\left({\bf y}_{ih} \mid {\bf b}_{ih,q}, {\bf X}_{ih}\right),
\eeq
where ${\bf y}_i$ denotes the observed value of ${\b Y}_i$ and ${\bf b}_{i,q} = ({\bf b}_{i1,q}, \dots, {\bf b}_{iH,q})$.
To complete the model structure, we further assume that the conditional response density is defined by an Asymmetric Least Informative Distribution \citep[ALID -][]{Bianchial:14, AlfoRanalliSalvati2016}. That is, for a given $q \in (0,1)$, we adopt the following density for the observed responses:
\beq \label{ALID}
Y_{ith} \mid \tbf b_{ih, q} \sim \frac{1}{B_q(\sigma_q)} \exp\{-\rho_q[y_{ith} -MQ_q(y_{ith} \mid \tbf x_{ith},\tbf b_{ih, q}; \psi)]\},
\eeq
where $\rho_q(\cdot)$ is the Huber loss function,  $B_q(\cdot)$ is a normalizing constant that ensures the density integrates to one, and $\sigma_q$ is a M-quantile-specific scale parameter. See \cite{Bianchial:14} for  details and properties of the ALID density. Note that the ALID is just a working density model that allows us to cast standard estimation of M-quantile regression models into a maximum likelihood (ML) context. This approach is similar to the one used in quantile regression modelling, where the Asymmetric Laplace Distribution \citep[ALD -][]{YuMoyeed2001} is frequently considered for similar purposes. 

For the $h$-th outcome and a fixed $q \in (0,1)$, we assume that the M-quantile of the (conditional) density in equation (\ref{ALID}) is described by the following linear model:
\beq 
MQ_q(y_{ith} \mid \tbf x_{ith}, \tbf b_{ih,q}; \psi)=\tbf x_{ith}^\prime \bs\beta_{h,q}+\tbf w_{ith}^\prime \tbf{b}_{ih,q}, \quad  h=1,\ldots,H,\label{eq:MQ}
\eeq
where $\psi$ denotes the influence function corresponding to the chosen loss function and $\b \beta_{h,q}$ is a $p$-dimensional vector of fixed effects for the $h$-th response. Random parameters $\tbf b_{ih,q}$ are associated with a subset of the design vector $\tbf x_{ith}$, that is $\tbf w_{ith}$. As we outlined before, these are added to the linear predictor to account for unobserved heterogeneity. 

Since the random parameters are unobserved, a way to estimate the model in equation (\ref{eq:MQ}) is based on treating $\tbf b_{ih,q}$ as \emph{nuisance} parameters with a density, $f_{b,q}(\cdot \mid \b \Omega_q)$, where $\b \Omega_q$ denotes a (possibly M-quantile dependent) covariance matrix. In this respect, the individual contribution to the \emph{observed data} log-likelihood for the $q$-th M-quantile can be obtained by integrating the conditional density in equation (\ref{mulconddens}) with respect to the random parameter distribution (conditional on the observed covariates):
\beq \label{indcontloglik}
\ell_{i,q}\left( \cdot\right)= \log \left\{\int_{\mathcal{B}}f_{q}\left({\bf y}_{i} \mid {\bf b}_{i,q}, {\bf X}_{i}\right)f_{b,q}({\bf b}_{i,q} \mid {\bf X}_{i}, \b \Omega_q){\rm d}{\bf b}_{i,q} \right\}.
\eeq

Therefore,  the log-likelihood function is defined by the following expression:
\beq \label{loglik}
\ell_{q}\left(\cdot \right)= \sum_{i=1}^{n}\ell_{i,q}\left(\cdot\right).
\eeq
As it may be noticed by looking at equation (\ref{indcontloglik}), the integral is defined with respect to the distribution of the random parameters, conditional on the observed covariates. As we will discuss in the following section, such a dependence should be properly handled.

\section{Handling potential endogeneity}\label{endog}
When working with random parameter models, a standard assumption is that observed covariates are uncorrelated with (in general independent of) sources of unobserved heterogeneity that describe the dependence between responses recorded on the same sample unit. Formally, this translates into the following equality: 
\[
f_{b,q}({\bf b}_{i,q} \mid {\bf X}_{i}, \b \Phi_q)=f_{b,q}({\bf b}_{i,q} \mid \b \Phi_q),
\]
leading to the assumption of \emph{strict exogeneity} of observed covariates:
\[
{\rm E}({\bf b}_{ih,q} \mid {\bf X}_{ih}) = {\rm E}({\bf b}_{ih,q}) = {\bf 0}. 
\]
In turn, the above expression implies \emph{weak exogeneity} of observed covariates, that is:
\[
{\rm E}({\bf b}_{ih,q}^{\prime} {\bf X}_{ih})={\rm cov}({\bf b}_{ih,q}, {\bf X}_{ih})={\bf 0}.
\]
Such an assumption can not be properly tested and is implicitly based on the orthogonality between  observed and omitted covariates. As it is clear, also this latter assumption is difficult to be tested. 
The above hypotheses are usually required for the model parameters to be identified, since the vector of variables ${\bf w}_{ith}$ is usually a subset of the design vector ${\bf x}_{ith}$, and the corresponding effects, ${\bf b}_{ih,q}$, are assumed to be zero mean deviations from the corresponding elements in $\bs\beta_{h,q}$. However, if this assumption is not fulfilled,  the ML estimate of $\bbeta_{h,q}$ will represent both the effect of $\tbf X_{ih}$ and the influence of the omitted covariates. As a result, this will represent an inconsistent estimate of the \emph{true} parameter vector, resulting in a non-interpretable mixture of effects \citep[see e.g.][]{Amemiya1984}. 

Here, we propose to handle the potential dependence of the random parameters on the observed covariates by adopting the following parametrisation:
\[
{\bf b}_{ih,q}=\sum_{t=1}^{T_i} {\bf \Lambda}_{th,q} \tbf x_{ith}+ {\bf b}_{ih,q}^{*},
\]
where ${\bf \Lambda}_{th,q}$ is an $s\times p$ matrix of coefficients. 
In the equation above, $\sum_{t=1}^{T_i} {\bf \Lambda}_{th,q} \tbf x_{ith}$ represents the best linear predictor of ${\bf b}_{ih,q}$ based on $\tbf x_{ith}$, while ${\bf b}_{ih,q}^{*}$ denotes a \textit{residual} latent effect which is (linearly) free from $\tbf X_{ih}$. If we further assume that ${\bf \Lambda}_{th,q}={\bf \Lambda}_{h,q}, \forall t=1,\dots, T_i $, we obtain the following \emph{auxiliary} equation:
\[
{\bf b}_{ih,q}={\rm E}({\bf b}_{ih,q} \mid {\bf X}_{ih}) + {\bf b}_{ih,q}^{*}= {\bf \Lambda}_{h,q} \sum_{t=1}^{T_i} \tbf x_{ith}+ {\bf b}_{ih,q}^{*}={\bf \Lambda}_{h,q} T_i\bar{\tbf x}_{ih}+ {\bf b}_{ih,q}^{*},
\]
where $\bar{\tbf x}_{ih} = T_i^{-1} \sum_t {\tbf x}_{ith} $. Using such a parametrisation, the M-quantile regression model in equation (\ref{eq:MQ}) becomes:

\begin{align*}
MQ_q(y_{ith} \mid \tbf x_{ith}, \tbf b_{ih,q}; \psi)
&=\tbf x_{ith}^\prime \bs\beta_{h,q}+\tbf w_{ith}^\prime \tbf{b}_{ih,q}
\nonumber \\ 
&= \tbf x_{ith}^\prime \bs\beta_{h,q} + \tbf w_{ith}^\prime \left[{\bf \Lambda}_{h,q}T_i\bar{\tbf x}_{ih}+ {\bf b}_{ih,q}^{*}\right]
\nonumber \\ 
&= \tbf x_{ith}^\prime \bs\beta_{h,q}+ \left(\tbf w_{ith} \otimes T_i \bar{\tbf x}_{ih}\right)^\prime {\blambda_{h,q}}+ {\bf w}_{ith}^{\prime}{\bf b}_{ih,q}^{*},
\label{eq:M2}
\end{align*}
where $\blambda_{h,q}={\rm vec}({\bf \Lambda}_{h,q})$.

This approach dates back at least to the \emph{auxiliary regression} method proposed by \cite{Mundlak1978} and subsequently extended by \cite{Chamberlain1980,Chamberlain1984} who defined the so-called \emph{correlated} random effect estimator. The term $\bar{\tbf x}_{ih}$ is used as an (informal) \emph{instrument} that renders the {residual} latent effects ${\bf b}_{ih,q}^{*}$ free from $\tbf X_{ih}$. Obviously, this is a specific parametrisation which may not be the \emph{optimal} one; however, given that the coefficient ${\bf \Lambda}_{h,q}$ varies with $q$, it may be considered as general enough to handle several forms of dependence of the random parameters on the observed covariates. In the following, we will refer to the re-parametrised random parameters as ${\bf b}_{ih,q}$ dropping the $*$ to simplify notation. 

It is worth to notice that a similar re-parametrisation was used by \cite{WuCarroll1988}, \cite{WuBailey1989}, \cite{FollmannWu1995} who referred to it as the \emph{approximate conditional model} in missing data modelling. A strong connection can also be found with the solutions to the \emph{initial conditions} problem proposed by \cite{AitkinAlfo1998}, \cite{FotohuiDavies1997} and \cite{Wooldridge2005}; see \cite{RabeSkrondal2014} for a review of the topic.
Also, a similar approach was proposed, in the quantile regression context, by \cite{AbrevayaDahl2008}. In this case, available data come from a longitudinal study with only two measures per statistical unit.
As the model may depend on individual-specific latent effects and may, therefore, have a complex form, the authors suggested to fit a linear quantile regression model as a reliable approximation to the \emph{true}, but unknown, quantile function, and to distinguish a \emph{direct} and an \emph{indirect} effect of the covariates on the observed responses.

\section{Estimation of the random effect distribution}\label{semip}
As mentioned earlier, to enhance model flexibility and avoid unverifiable assumptions on the random parameter distribution, we adopt the approach developed by \citet{AlfoRanalliSalvati2016}, who introduced a finite mixture of M-quantile regression models. This proposal is more in line with the robust approach of M-quantile regression, as it avoids the assumption of Gaussian random effects, that is usually adopted in the longitudinal setting.
In particular, we assume that random parameters follow a discrete distribution defined over the (outcome- and M-quantile-specific) support set $\{\b \zeta_{1h, q}, \dots, \b \zeta_{Kh, q}\}$ with masses $\pi_{k,q} = \Pr({\bf b}_{ih,q}=\b \zeta_{kh, q}) \geq 0, \forall k = 1, \dots, K$, subject to $\sum_k \pi_{k,q} =1$. 
Using such a specification, locations $\{\b \zeta_{1h, q}, \dots, \b \zeta_{Kh, q}\}$ are associated across outcomes thanks to the common prior distribution. 
To be more specific, the outcome-specific distribution is given by
\[
\tbf b_{ih,q} \sim \sum_{k} {\pi_{k,q}} \delta(\b \zeta_{kh,q}),\quad h=1,\ldots,H,
\]
while the joint distribution of all individual-specific random parameters can be specified according to
\[
\tbf b_{i,q} \sim \sum_{k} {\pi_{k,q}}\delta(\b \zeta_{k1,q},\dots,\b \zeta_{kH,q}).
\]
In the above expressions, $\delta(\theta)$ denotes an indicator function putting a unit mass at $\theta$. Such an approach may be linked to the NonParametric Maximum Likelihood (NPML) estimate of the mixing distribution $f_{b,q}(\cdot)$; see \cite{Aitkin1999}. Conditional on the membership to the $k$-th component of the finite mixture, the M-quantile regression model of order $q$ in equation \eqref{eq:MQ} can be written as follows:
\beq
MQ_q(y_{ith} \mid \tbf x_{ith}, \b \zeta_{kh,q};\psi)=\tbf x_{ith}^\prime \bs\beta_{h,q}+ \left(\tbf w_{it} \otimes T_i
 \bar{\tbf x}_{ih}\right)^\prime {\blambda_{h,q}}+ \tbf w_{ith}^\prime \b \zeta_{kh,q}.\label{eq:MQ3}
\eeq
In the next section, we describe estimation of model parameters in a maximum likelihood framework.

\section{Inference on model parameters}\label{sec:4}
In this section, we present computational details of the EM algorithm we exploit to obtain ML parameter estimates and the procedure we use to compute the corresponding analytic standard errors. 

\subsection{ML estimation}
Based on the independence assumption within the same outcome (conditional on the individual- and outcome-specific random parameters), $\tbf b_{ih,q}$, and between outcomes (conditional on the vector of individual-specific random parameters), $\tbf b_{i,q}$, the joint conditional density for responses associated to a generic unit in $k$-th component of the finite mixture is given by 
\[
f_q({\bf y}_{i} \mid {\bs \beta}_{q}, \blambda_{q}, {\b \zeta}_{k,q}, {\b \sigma_{q}}) = 
\prod_{h= 1}^{H} \prod_{t = 1}^{T_i} f_q(y_{ith} \mid {\bs \beta}_{h,q}, \blambda_{h,q}, {\b \zeta}_{kh,q}, {\sigma_{h,q}}).
\]
In the expression above, ${\bs \beta}_{q},\;  \b \lambda_q, \; {\b \zeta}_{k,q}$, and ${\b \sigma_{q}}$ are vectors including all outcome- and M-quantile-specific parameters for varying $h = 1, \dots, H$. The dependence of the response distribution on such parameters is here explicitly shown to provide a clearer description of the estimation algorithm.

Let ${\bs\Phi}_{q}$ denote the \textit{global} set of model parameters for the $q$-th M-quantile, $q \in (0,1)$.  The observed data likelihood is defined by the following expression:
\begin{equation}
L({\bs\Phi}_{q}) = \prod_{i= 1}^n \sum_{k = 1}^K  f_q({\bf y}_{i} \mid {\bs \beta}_{q}, \blambda_{q}, {\b \zeta}_{k,q}, {\b \sigma_{q})\: \pi_{k,q}}.
\label{eq:lik2}
\end{equation}
Although estimates can be directly obtained by differentiating equation \eqref{eq:lik2}, an indirect approach based on the use of the EM algorithm \citep{Dempster1977} is simpler to implement. Let us start from the definition of the {\em complete data} log-likelihood for a given M-quantile $q$: 
\begin{eqnarray}
\ell_{c}(\bs\Phi_q)= \sum_{i=1}^{n}\sum_{k=1}^{K} z_{ik,q} \left\{\log \left[f_q({\bf y}_{i} \mid {\bs \beta}_{q}, \blambda_{q}, {\b \zeta}_{k,q}, \b \sigma_{q})
\right]+\log(\pi_{k,q})\right\},  \label{lcq}
\end{eqnarray}
where $z_{ik,q}$, $i=1,\dots,n$, $k=1,\dots,K$, denotes the indicator variable for the $i$-th individual in the $k$-th component of the finite mixture.

Starting from the equation above, parameter estimates are derived by alternating two separate steps. 
In the E-step, we compute the posterior expectation of the {complete data} log-likelihood in equation \eqref{lcq}, conditional on the observed data and the current value of model parameters, say ${\b \Phi_q^{(r-1)}}$. That is, at the $r$-th step of the EM algorithm, the indicator variables  in  the {complete data} log-likelihood are replaced by the corresponding conditional expectations: 
\[
\hat z_{ik,q}^{(r)} = \frac{\pi_{k,q}^{(r-1)} 
f_q({\bf y}_{i} \mid {\bs \beta}_{q}^{(r-1)}, \blambda_{q}^{(r-1)}, {\b \zeta}_{k,q}^{(r-1)}, \b \sigma_{q}^{(r-1)})}
{\sum_{g=1}^K \pi_{g, q}^{(r-1)} f_q({\bf y}_{i} \mid {\bs \beta}_{q}^{(r-1)}, \blambda_{q}^{(r-1)}, {\b \zeta}_{g,q}^{(r-1)}, \b \sigma_{q}^{(r-1)})
}.
\]
This leads to the following expected complete data log-likelihood: 
\begin{equation}
Q(\bs
 \Phi_q \mid {\b \Phi_q^{(r-1)}}) =\sum_{i=1}^{n}\sum_{k=1}^{K} z_{ik,q}^{(r)} \left\{\log\left[f_q({\bf y}_{i} \mid {\bs \beta}_{q}, \blambda_{q}, {\b \zeta}_{k,q}, \b \sigma_{q}) \right]+\log(\pi_{k,q})\right\}.  \label{eq:Qfnc}
\end{equation}
In the M-step, equation \eqref{eq:Qfnc} is maximized to derive updated estimates ${\b \Phi_q^{(r)}}$. 
Due to the separability of the parameter set, maximization can be partitioned into different sub-problems, consistently simplifying the computation. Closed form expressions are available for the mixture priors: 
\[
\hat \pi_{k,q} = \frac{\hat z_{ik,q}^{(r)}}{n}.
\]
On the other hand, longitudinal model parameters are estimated by a weighted version of the standard Iteratively Weighted Least Squares algorithm for M-quantile regression. 
It is worth noticing that a similar approach can be used when dealing with quantile regression. In this case, we simply replace the ALID function in equation \eqref{ALID} by the Asymmetric Laplace Distribution (ALD) by \cite{YuMoyeed2001}. Then, updated estimates for regression model parameters can be obtained by defining an appropriate linear programming algorithm,  which turns out to be a weighted version of the usual simplex-type algorithm for cross-sectional quantile regression.
For a detailed description of the algorithm, both for M-quantile and quantile regression, see \cite{AlfoRanalliSalvati2016}.  

To get the final estimate of $\bPhi_q$, the E- and the M- steps are repeatedly alternated until the difference between two subsequent likelihoods, or rather the norm $\|{\b \Phi_q^{(r)}} - {\b \Phi_q^{(r-1)}}\|$, is lower than a fixed constant $\varepsilon>0$.

\subsection{Computation of standard errors and model selection}\label{sec_se}
Estimation of the covariance matrix for $\hat{\b \Phi}_q$ can be done analytically by exploiting the Oakes' formula, see \cite{Oakes1999}.  According to the simulation results discussed in \cite{FriedlKauermann2000}, we decided to derive standard errors by using the \emph{sandwich} estimate, see \cite{Royall1986} and \cite{white1980}. 
Let $\hat{\b \Phi}_q$ denote the ML parameter estimate of $\bPhi_q$ and let  $\bPhi_q^\star$ denote the true parameter vector for a given M-quantile $q$. The sandwich estimate of the corresponding covariance matrix is computed by using the following expression:
\begin{equation}
\label{eq_sandwhich}
\widehat{\mbox{Cov}}(\hat{\b \Phi}_q)  = {\b J}(\hat{\b \Phi}_q )^{-1} {\b K}(\hat{\b \Phi}_q ) {\b J}(\hat{\b \Phi}_q )^{-1},
\end{equation}
where 
$$
{\b J}(\hat{\b \Phi}_q ) = - \left[\frac{\partial \ell(\b \Phi_q)}{ \partial \b \Phi_q \partial \b \Phi_q^\prime}\right]_{\b \Phi_q = \hat{\b \Phi}_q},
$$
represents the observed information matrix, while ${\b K}(\hat{\b \Phi})$ provides an estimate of the score covariance matrix
\begin{equation}
\label{eq_scoreCov}
{\b K}({\b \Phi^*_q}) =  \text{cov} \left[ \frac{\partial \ell(\b \Phi_q)}{\partial \b \Phi_q} \right]_{{\b \Phi_q } = {\b \Phi^*_q }}=\sum_i \text{cov} \left[ \frac{\partial \ell_i(\b \Phi_q)}{\partial \b \Phi_q} \right]_{{\b \Phi_q } = {\b \Phi^*_q }}.
\end{equation}

In practice, the quantity $\b {J}(\hat{\b \Phi }_q)$ in equation \eqref{eq_sandwhich} can be computed from the Oakes formula \citep{Oakes1999}: 
\begin{equation}
\b{J}({\hat{\b \Phi}}_q)=-\left\{\frac{\partial^2 Q({\bs  \Phi_q} 
 \mid \hat{\bs
 \Phi}_q)}{\partial {\bs \Phi_q} \partial {\bs  \Phi_q}^{\prime}}\right|_{{\bs  \Phi_q}=\hat{\bs \Phi}_q} + \left. \left. \frac{\partial^2 Q({\bs
 \Phi_q} \mid \hat{\bs
 \Phi}_q)}{\partial {\bs \Phi_q} \partial \hat{\bs
 \Phi}_q^{\prime}}\right|_{{\bs \Phi}_q=\hat{\bs \Phi}_q
}\right\}. \label{oakes}
\end{equation}
In this expression, the first term is directly obtained from the EM algorithm and denotes the expectation of the {complete-data} Hessian matrix, conditional on the observed data and the parameter estimates at convergence. On the other hand, the second term appearing in expression \eqref{oakes} denotes the first derivative of the expected complete-data score with respect to the current parameter estimates. This quantity is obtained by computing the numerical derivative of the score function, where only the posterior weights of the finite mixture are considered as a function of the parameters. 

Last, an estimate of the score covariance matrix in equation \eqref{eq_scoreCov}, ${\b K}(\hat{\b \Phi}_q)$, is given by 
$$
{\b K}(\hat{\b \Phi}_q ) = \sum_{i = 1}^n S_i(\hat{\b \Phi}_q)S{_i}(\hat{\b \Phi}_q)^\prime,
$$ 
with $S_i(\hat{\b \Phi}_q)$ being the individual score for the $i$-th individual, evaluated at $\hat{\b \Phi}_q$.
Standard errors for $\hat{\bPhi}_q$ are obtained by taking the square root of the diagonal elements of $\widehat{\mbox{Cov}}(\hat{\b \Phi}_q)$.

It is worth noticing that, if we move from the M-quantile to the quantile regression framework by using an ALD in place of the ALID, standard errors for parameter estimates cannot be estimated analytically any longer. This is due to non-differentiability of the likelihood function. In that framework, nonparametric bootstrap represents the usual strategy to obtain a measure of dispersion for parameter estimates \citep[see e.g.][]{AlfoRanalliSalvati2016}. 

As it is frequently done in the mixture model framework, for a fixed $q \in (0,1)$, the algorithm is run for fixed $K$; once it reaches convergence, $K$ is increased to $K+1$ and the algorithm is run again. A formal comparison between models corresponding to different choices for $K$ can be performed using penalized likelihood criteria, e.g. AIC \citep{Akaike1973} or BIC \citep{Schwarz1978}. As it is typically done when dealing with latent variables, we suggest to consider a multi-start strategy to avoid the algorithm being trapped in local maxima solutions. That is, for fixed $K$, we suggest to run the algorithm starting from different initial values and retain the best solution in terms of maximized log-likelihood.

\section{Analysis of the Millennium Cohort Study data}\label{sec:7}

In this section, we analyse data from the Millennium Cohort study via the M-quantile regression model for multivariate longitudinal responses discussed so far. We aim at understanding how available covariates, especially those associated with neighbourhood and family risk factors, influence internalizing and externalizing depression symptoms in UK children, {measured via the i-SDQ and e-SDQ indexes, respectively.}
A similar analysis was proposed by \cite{Tzavidis2016}, where two separate univariate, mixed, M-quantile regression models for i-SDQ and e-SDQ were estimated. Here, we extend this analysis to provide a deeper description of the MCS data. In particular, we are interested in analysing  how the social and economic conditions \textit{jointly} influence the evolution of the responses over time. Also, we aim at controlling for sources of unobserved heterogeneity that determine the dependence both between and within individual profiles, adopting a nonparametric specification for the random parameters distribution. Last, we aim at distinguishing direct and indirect effects (mediated by the random parameters) on the response to obtain a clearer interpretation of the results.

For a given $q \in (0,1)$, we consider the following model specification: 
\begin{equation}\label{eq:MMq_application}
MQ_q(y_{ith} \mid \tbf x_{ith},\b  \zeta_{kh,q}; \psi) = \zeta_{kh,q} + \tbf x_{1ith}^\prime \b \beta_{1h,q} + (\tbf x_{2ith}-\bar{\tbf x}_{2ih}) ^\prime \b \beta_{2h,q}  + \bar{\tbf x}_{2ih}^\prime \b \lambda_{h,q}+ \tbf x_{3ih}^\prime \b \beta_{3h,q},
\end{equation}
for $h = 1,2, \, t = 1, \dots, 3$, and $i = 1, \dots, n,$ with $n = 9,$021.

The vector $\tbf x_{1ith} = \tbf x_{1it}$ includes the variable Age (which is centred around the overall mean) and its quadratic transform ($\mbox{Age}^2$). On the other hand, $\tbf x_{i2th} = \tbf x_{i2t}$ denotes the vector of time-varying covariates which are centred around their individual mean, $\bar{\tbf x}_{2ih} = \bar{\tbf x}_{2i}$. This set includes the variables $\mbox{ALE}_{11}$, $\mbox{SED}_4$, Kessm, and the $\mbox{IMD}$. Last, $\tbf x_{3ih} = \tbf x_{3i} $ denotes the vector of time-constant covariates and includes maternal education (reference = no qualification), ethnicity (reference = non-white), gender (reference = female) and the stratification variable (reference = advantaged stratum). 
The vectors $\b \beta_{1h,q}$ and $\b \beta_{2h,q}$ are meant to represent direct effects of observed covariates and are often referred to as \textit{within} effects. On the other hand, $\b \lambda_{h,q}$ represents the indirect effect mediated by random parameters and, together with $\b \beta_{3h,q}$, are often referred to as \textit{between} effects. These two quantities are associated with ``age/time'' (within) and ``cohort'' (between) effects, respectively, with the former entailing individual unobserved dynamics, and the latter corresponding to time-constant heterogeneity. 

In this application, individual-specific features and behaviours may be likely associated to demographic and socio-economic conditions which, in turn, affect the value of the observed covariates. To solve this issue, we follow the approach detailed in Section \ref{endog} and consider the M-quantile regression model in equation \eqref{eq:MMq_application}.
Also, we focused on the right tail of the response distribution, which is associated with more severe emotional and behavioral problems. For this purpose, we estimated the M-quantile regression model for $q = \{0.50, 0.75, 0.90\}$ with a varying number of mixture components ($K = 2, \dots, 15$). 

In order to avoid that the algorithm remains trapped in local maxima, we considered a multi-start strategy, for each value of $K$. A first deterministic solution was obtained by fixing component probabilities $\pi_{k,q} = 1/K, k = 1, \dots,K$, while values for fixed parameters in the longitudinal data model were derived from a univariate homogeneous linear model. 
Component-specific random intercepts were obtained by adding $K$ Gaussian quadrature locations to the corresponding (fixed) effects from the linear model. For each value $K$, we considered $d(K-1)$ random starting solutions, with $d = 3$. These were obtained from the deterministic ones, by randomly perturbing model parameters. The solution corresponding to the highest log-likelihood value was retained as the optimal one. 
To avoid spurious solutions or estimates lying on the boundary of the parameter space, and to facilitate the interpretation of the results, for each level $q$, we performed the search for the optimal $K$ value using the BIC index among the solutions fulfilling $\pi_{k,q} \geq 0.01, \forall k = 1, \dots, K$. This led us to select a model with $K=\{10, 7, 3\}$ for $q = \{0.50, 0.75, 0.90\}$, respectively. 

Figure \ref{fig_uVal_SDQ} shows the estimated cumulative density function of the random parameters for the internalizing (left) and the externalizing (right) scores, at different M-quantiles. 
\begin{figure}
\caption{The Millennium Cohort Study data. Estimated cumulative density function of the discrete random parameters for i-SDQ (left panel) and e-SDQ (right panel) at different M-quantile levels.}
\centering
\vspace{2mm}
\includegraphics[scale=0.25]{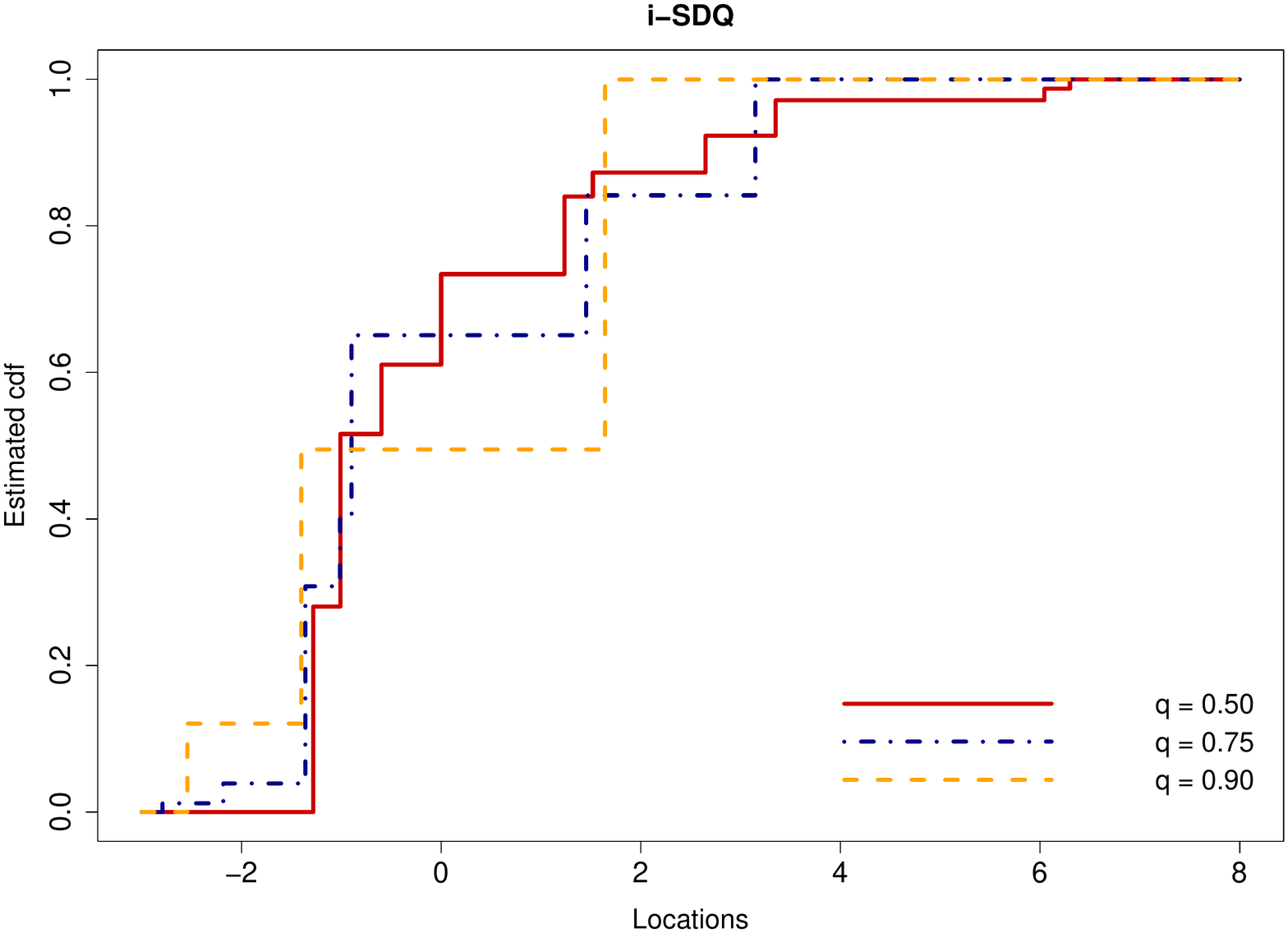} 
\includegraphics[scale=0.25]{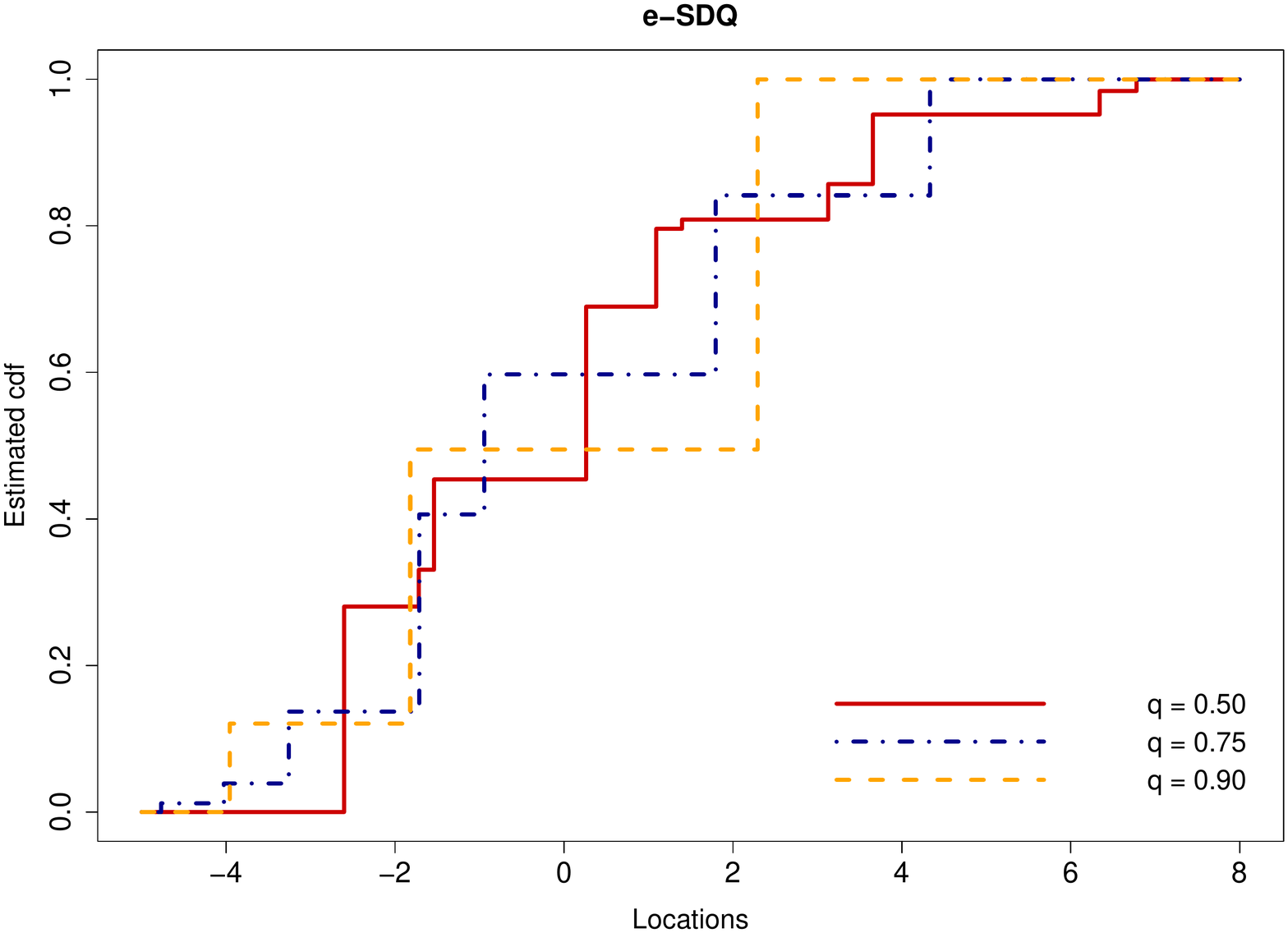} 
\label{fig_uVal_SDQ}
\end{figure}
As it can be observed by looking at the spread of the estimated locations, e-SDQ  shows a higher variability than i-SDQ. Also, we notice that the probability of observing higher locations increases as we move from $q = 0.50$ to $q = 0.90$. Last, the estimated distribution of both set of random effects (related either to i-SDQ or e-SDQ) is quite far from being symmetric and unimodal. Such a finding renders the proposed semi-parametric approach an interesting solution for the analysis of the SDQ data where, clearly, the standard assumption of Gaussian distributed random parameters does not seem to hold.

Estimates for the fixed parameters in the model for the two SDQ outcomes and the three M-quantile levels $q= \{0.50, 0.75, 0.90\}$, are reported in Tables \ref{tab_iSDQ} and \ref{tab_eSDQ}, respectively. Standard errors of the estimates were obtained according to the procedure detailed in Section \ref{sec_se}. Focusing on the i-SDQ scores (see Table \ref{tab_iSDQ}), we observe that age plays a minor (although significant) role in determining the evolution of the responses over time for all the analysed M-quantiles. Neighbourhood and family risk factors play a central role in explaining both between children heterogeneity and the evolution of the i-SDQ scores over time. 

Adverse life events, family poverty measured by $\mbox{SED}_4$ and maternal depression are  all positively associated with i-SDQ: the worse the socio-economic conditions of children, the higher the scores. Such an effect is stronger when moving from $q = 0.50$ to $q= 0.90$. In particular, we notice that differences between units in the sample can be explained in terms of these variables (positive and significant effect for the corresponding individual means), with an effect that increases for higher values of $q$. On the other hand, looking at the standard errors for the estimated effects of $(\mbox{SED}_4 -\overline{\mbox{SED}}_4)$, $(\mbox{ALE}_{11} -\overline{\mbox{ALE}}_{11})$, and $(\mbox{Kessm} - \overline{\mbox{Kessm}})$, we conclude that the evolution of the i-SDQ scores over time is only influenced by the evolution of adverse life events ($\mbox{ALE}_11-\overline{\mbox{ALE}}_{11}$), with an effect that becomes stronger at the right tail of the distribution. 

As far as the IMD variable is concerned, we observe a significant effect for the individual mean variable, $\overline{\mbox{IMD}}$, on the highest M-quantiles only, that is $q = \{0.75, 0.90\}$. The negative sign of these parameters highlights that children living in less deprived areas (higher IMD) show lower internalizing problems and this effect is stronger for higher scores. In particular, children with higher mean IMD value, that is, usually living in less deprived areas, tend to have lower scores.

Furthermore, higher i-SDQ scores are observed for black children and male children (positive estimate for Male at $q= \{0.75, 0.90\}$ and negative effect for White), at all M-quantiles. More severe internalizing scores are less likely in the presence of higher mother's educational level (negative effect for GCSE and Degree) and the absolute size of these effects becomes stronger as we move from the center to the right tail of the distribution. Last, the stratification variable has a non-significant effect after controlling for the remaining covariates.

\begin{table}
\caption{The Millennium Cohort Study: internalizing scores (i-SDQ). Fixed parameter estimates at different M-quantiles.}
\vspace{2mm}
\centering
\begin{tabular}{lrrrrrrrrrr}
\toprule
\toprule
& \multicolumn{1}{c}{$q = 0.50$}& \multicolumn{1}{c}{$q = 0.75$} & \multicolumn{1}{c}{$q = 0.90$}\\
\midrule

Age	&	-0.021	(0.008)	&	-0.010	(0.011)	&	0.037	(0.015)	\\
$\mbox{Age}^{2}$	&	0.068	(0.005)	&	0.085	(0.006)	&	0.090	(0.010)	\\
$\overline{\mbox{ALE}}_{11}$	&	0.098	(0.022)	&	0.193	(0.038)	&	0.368	(0.057)	\\
$(\mbox{ALE}_{11} -\overline{\mbox{ALE}}_{11})$	&	0.056	(0.017)	&	0.077	(0.022)	&	0.104	(0.032)	\\
$\overline{\mbox{SED}}_4$		&	0.133	(0.039)	&	0.116	(0.055)	&	0.206	(0.078)	\\
$(\mbox{SED}_4-\overline{\mbox{SED}}_4)$		&	-0.041	(0.031)	&	-0.033	(0.042)	&	-0.049	(0.058)	\\
$\overline{\mbox{Kessm}}$	&	0.172	(0.009)	&	0.242	(0.014)	&	0.354	(0.020)	\\
$(\mbox{Kessm} -\overline{\mbox{Kessm}})$		&	0.083	(0.009)	&	0.100	(0.012)	&	0.127	(0.017)	\\
Degree	&	-0.665	(0.071)	&	-0.784	(0.121)	&	-1.047	(0.138)	\\
Gcse	&	-0.410	(0.068)	&	-0.483	(0.112)	&	-0.626	(0.131)	\\
White	&	-0.304	(0.057)	&	-0.337	(0.123)	&	-0.417	(0.130)	\\
Male	&	0.044	(0.037)	&	0.168	(0.050)	&	0.349	(0.089)	\\
$\overline{\mbox{IMD}}$ 	&	-0.025	(0.008)	&	-0.048	(0.015)	&	-0.087	(0.020)	\\
$(\mbox{IMD} -\overline{\mbox{IMD}})$	&	-0.005	(0.018)	&	-0.012	(0.024)	&	-0.002	(0.037)	\\
Ethnic St.	&	0.163	(0.095)	&	0.225	(0.126)	&	0.137	(0.157)	\\
Disadv St.	&	0.039	(0.054)	&	0.055	(0.066)	&	-0.017	(0.113)	\\
	
$\sigma_{h,q}$	&	\multicolumn{1}{c}{1.720}		&	\multicolumn{1}{c}{1.730}		&	\multicolumn{1}{c}{1.700}		\\
\bottomrule
\bottomrule
\end{tabular}
\label{tab_iSDQ}
\end{table}

When looking at the results for the behavioral problems (e-SDQ scores) reported in Table \ref{tab_eSDQ}, we first observe that, for all M-quantiles, e-SDQ appears to reduce until children are about $5$ years old (the overall mean age) and remain quite stable afterwards (see the estimated coefficients for Age and $\mbox{Age}^2$). As regards the covariates measuring neighbourhood and family risk factors, we draw conclusions which are similar to those for e-SDQ. 
However, in this case, parameter estimates have generally a higher magnitude than before, suggesting a stronger impact of such covariates on the response. The individual mean $\overline{\mbox{ALE}}_{11}$ has a positive and significant effect on the response at all M-quantiles we considered, while $\overline{\mbox{ALE}}_{11}$ and $({ALE}_{11} -\overline{\mbox{ALE}}_{11})$ positively influence the response at $q = \{0.75, 0.90\}$, only. 
As before, maternal depression, measured both in terms of individual means and in terms of deviations from the latter, has a positive effect on the outcome with a magnitude that becomes higher when moving towards the right tail of the distribution. Also in this case, the effect associated with the mean covariate (between) appears to be higher, thus pointing to the quality of the environment rather than to its change. Higher values for the individual IMD averages correspond to lower e-SDQ scores for $q = 0.75$ and $q = 0.90$, while the effect is negligible for the median.
Furthermore, the effect of gender and race is more evident for the externalizing scores when compared to the internalizing ones: black children and male children exhibit  higher scores, especially for higher M-quantiles. As before, the stratification variable has a negligible impact on the response variable after controlling for other covariates.

\begin{table}
\caption{The Millennium Cohort Study: externalizing scores (e-SDQ). Fixed parameter estimates at different M-quantiles.}
\vspace{2mm}
\centering
\begin{tabular}{lrrrrrrrrrr}
\toprule
\toprule
& \multicolumn{1}{c}{$q = 0.50$}& \multicolumn{1}{c}{$q = 0.75$} & \multicolumn{1}{c}{$q = 0.90$}\\
\midrule

Age	&	-0.448	(0.010)	&	-0.466	(0.013)	&	-0.461	(0.019)	\\
$\mbox{Age}^2$	&	0.214	(0.007)	&	0.235	(0.008)	&	0.250	(0.012)	\\
$\overline{\mbox{ALE}}_{11}$	&	0.187	(0.040)	&	0.323	(0.060)	&	0.505	(0.079)	\\
$(\mbox{ALE}_{11}- \overline{\mbox{ALE}}_{11})$		&	0.089	(0.022)	&	0.096	(0.027)	&	0.096	(0.038)	\\
$\overline{\mbox{SED}}_4$		&	0.176	(0.057)	&	0.234	(0.071)	&	0.345	(0.105)	\\
$(\mbox{SED}_4 - \overline{\mbox{SED}}_4)$  	&	-0.006	(0.041)	&	0.004	(0.046)	&	0.008	(0.069)	\\
$\overline{\mbox{Kessm}}$  	&	0.225	(0.014)	&	0.261	(0.023)	&	0.357	(0.026)	\\
$(\mbox{Kessm}-\overline{\mbox{Kessm}})$	&	0.112	(0.011)	&	0.126	(0.012)	&	0.160	(0.018)	\\
Degree	&	-1.171	(0.131)	&	-1.398	(0.175)	&	-1.654	(0.215)	\\
Gcse	&	-0.472	(0.121)	&	-0.596	(0.153)	&	-0.751	(0.185)	\\
White	&	0.169	(0.093)	&	0.418	(0.221)	&	0.372	(0.239)	\\
Male	&	0.753	(0.057)	&	0.968	(0.096)	&	1.252	(0.137)	\\
$\overline{\mbox{IMD}}$	&	-0.042	(0.014)	&	-0.049	(0.023)	&	-0.071	(0.033)	\\
$(\mbox{IMD} - \overline{\mbox{IMD}})$ 	&	-0.030	(0.025)	&	-0.032	(0.030)	&	-0.026	(0.044)	\\
Ethnic St.	&	-0.080	(0.115)	&	-0.052	(0.252)	&	-0.177	(0.246)	\\
Disadv St.	&	0.058	(0.086)	&	0.184	(0.115)	&	0.251	(0.178)	\\
$\sigma_{h,q}$	&	\multicolumn{1}{c}{2.510}		&	\multicolumn{1}{c}{2.540}		&	\multicolumn{1}{c}{2.400}		\\
\bottomrule
\bottomrule
\end{tabular}
\label{tab_eSDQ}
\end{table}

{The estimated regression coefficients are consistent with those obtained by \cite{Tzavidis2016} for both SDQ scores.
However, by adopting our model specification, we are able to distinguish ``between'' and ``within'' effects of the observed covariates on the two outcomes of interest and, therefore, provide a clearer interpretation of the effect of social and economic conditions on children's depression symptoms.

In the last row of Tables \ref{tab_iSDQ}-\ref{tab_eSDQ}, we report the estimated standard deviation of individual-specific random parameters. These are directly obtained from the estimated locations and masses of the discrete distribution reported in Figure \ref{fig_uVal_SDQ}. As it can be observed, comparing the results for the internalizing and the externalizing scores, a higher variability for the random parameters in the latter model is present. No significant differences are observed when moving from $q = 0.50$ to $q = 0.90$, thus suggesting that individual-specific sources of unobserved heterogeneity similarly influence the center and the right tail of the responses' distribution. This latter finding is not in line with that by \cite{Tzavidis2016} and this may be possibly due to the semi-parametric approach we use in this context. 
}

\subsection{Complete case analysis}

In this section, we provide results from a complete case analysis based on  those children in the MCS who were observed at all measurement occasions only. As highlighted in Section \ref{MCS_study}, complete longitudinal sequences ($T_i=3$) are available for $ 5,342$ sample units ($59.22\%$ of the whole sample). 
Such an analysis represents an essential step for validating the results presented in so far, which are all based on the assumption of independence between observed data and the missing data mechanism generating the unobserved responses. Should the MAR mechanism hold, inferential conclusions based on complete longitudinal sequences only would not change. 

Following a similar strategy to that described in Section \ref{sec:7}, we estimated the M-quantile regression model for the joint analysis of internalizing and externalizing SDQ scores for $q = \{0.50, 0.75, 0.90\}$ with $K = 1, \dots, 15$. We considered the same multi-start approach described before to avoid local maxima and retained the solution corresponding to the maximum value of the log-likelihood function. As before, we considered the BIC index to identify the optimal number of mixture components, focusing the attention on those solutions ensuring $\pi_{k,q}>0.01, \forall k=1, \dots, K$. This led us to select, again, a model with $K = \{10,7,3\}$ for $q = \{0.50, 0.75, 0.90\}$, respectively.
We report in Tables \ref{tab_iSDQ_complete}-\ref{tab_eSDQ_complete} estimates for the fixed parameters in the model for i-SDQ and e-SDQ scores, respectively, together with the corresponding analytic standard errors (in brackets).

\begin{table}
\caption{Complete case analysis - The Millennium Cohort Study: internalizing scores. Fixed parameter estimates at different M-quantiles.}
\vspace{2mm}
\centering
\begin{tabular}{lrrrrrrrrrr}
\toprule
\toprule
& \multicolumn{1}{c}{$q = 0.50$}& \multicolumn{1}{c}{$q = 0.75$} & \multicolumn{1}{c}{$q = 0.90$}\\
\midrule

Age	&	-0.021	(0.008)	&	-0.018	(0.011)	&	0.022	(0.017)	\\
$\mbox{Age}^2$	&	0.069	(0.005)	&	0.090	(0.007)	&	0.101	(0.011)	\\
$\overline{\mbox{ALE}}_{11}$	&	0.084	(0.028)	&	0.190	(0.047)	&	0.469	(0.112)	\\
$(\mbox{ALE}_{11} - \overline{\mbox{ALE}}_{11})$	&	0.056	(0.018)	&	0.074	(0.024)	&	0.103	(0.036)	\\
$\overline{\mbox{SED}}_{4}$	&	0.143	(0.050)	&	0.102	(0.071)	&	0.190	(0.108)	\\
$(\mbox{SED}_{4} - \overline{\mbox{SED}}_{4})$	&	-0.009	(0.037)	&	0.014	(0.049)	&	0.028	(0.075)	\\
$\overline{\mbox{Kessm}}$		&	0.159	(0.011)	&	0.231	(0.017)	&	0.349	(0.027)	\\
$(\mbox{Kesmm} - \overline{\mbox{Kessm}})$	&	0.090	(0.010)	&	0.107	(0.014)	&	0.132	(0.019)	\\
Degree	&	-0.516	(0.095)	&	-0.410	(0.149)	&	-0.635	(0.229)	\\
Gcse	&	-0.293	(0.092)	&	-0.177	(0.148)	&	-0.292	(0.222)	\\
White	&	-0.245	(0.072)	&	-0.265	(0.127)	&	-0.280	(0.202)	\\
Male	&	0.038	(0.036)	&	0.137	(0.060)	&	0.330	(0.160)	\\
$\overline{\mbox{IMD}}$	&	-0.017	(0.009)	&	-0.042	(0.015)	&	-0.077	(0.029)	\\
$(\mbox{IMD} - \overline{\mbox{IMD}})$		&	-0.012	(0.020)	&	-0.020	(0.027)	&	-0.007	(0.043)	\\
Ethnic St.	&	0.156	(0.095)	&	0.182	(0.145)	&	0.162	(0.217)	\\
Disadv St.	&	0.033	(0.049)	&	0.032	(0.076)	&	-0.070	(0.137)	\\

$\sigma_{h,q}$	&	\multicolumn{1}{c}{1.511}		&	\multicolumn{1}{c}{1.553}		&	\multicolumn{1}{c}{1.565}		\\
\bottomrule
\bottomrule
\end{tabular}
\label{tab_iSDQ_complete}
\end{table}

\begin{table}
\caption{Complete case analysis - The Millennium Cohort Study: externalizing scores. Fixed parameter estimates at different M-quantiles.}
\vspace{2mm}
\centering
\begin{tabular}{lrrrrrrrrrr}
\toprule
\toprule
& \multicolumn{1}{c}{$q = 0.50$}& \multicolumn{1}{c}{$q = 0.75$} & \multicolumn{1}{c}{$q = 0.90$}\\
\midrule

Age	&	-0.449	(0.011)	&	-0.473	(0.013)	&	-0.465	(0.021)	\\
$\mbox{Age}^2$	&	0.212	(0.007)	&	0.237	(0.008)	&	0.256	(0.015)	\\
$\overline{\mbox{ALE}}_{11}$	&	0.131	(0.050)	&	0.310	(0.079)	&	0.595	(0.166)	\\
$(\mbox{ALE}_{11} - \overline{\mbox{ALE}}_{11})$	&	0.097	(0.024)	&	0.101	(0.030)	&	0.096	(0.044)	\\
$\overline{\mbox{SED}}_{4}$	&	0.226	(0.076)	&	0.251	(0.093)	&	0.444	(0.202)	\\
$(\mbox{SED}_{4} - \overline{\mbox{SED}}_{4})$	&	0.011	(0.049)	&	0.016	(0.056)	&	0.013	(0.101)	\\
$\overline{\mbox{Kessm}}$	&	0.233	(0.017)	&	0.253	(0.027)	&	0.378	(0.041)	\\
$(\mbox{Kesmm} - \overline{\mbox{Kessm}})$		&	0.116	(0.012)	&	0.128	(0.014)	&	0.171	(0.023)	\\
Degree	&	-1.037	(0.199)	&	-1.185	(0.238)	&	-1.427	(0.344)	\\
Gcse	&	-0.319	(0.194)	&	-0.411	(0.229)	&	-0.588	(0.329)	\\
White	&	0.157	(0.120)	&	0.464	(0.243)	&	0.413	(0.341)	\\
Male	&	0.723	(0.065)	&	0.914	(0.108)	&	1.304	(0.263)	\\
$\overline{\mbox{IMD}}$	&	-0.017	(0.016)	&	-0.020	(0.024)	&	-0.040	(0.051)	\\
$(\mbox{IMD} - \overline{\mbox{IMD}})$		&	-0.018	(0.027)	&	-0.025	(0.031)	&	-0.011	(0.051)	\\
Ethnic St.	&	0.036	(0.156)	&	0.038	(0.234)	&	-0.040	(0.321)	\\
Disadv St.	&	0.106	(0.091)	&	0.208	(0.128)	&	0.229	(0.213)	\\
$\sigma_{h,q}$	&	\multicolumn{1}{c}{2.407}		&	\multicolumn{1}{c}{2.508}		&	\multicolumn{1}{c}{2.334}		\\
\bottomrule
\bottomrule
\end{tabular}
\label{tab_eSDQ_complete}
\end{table}

When focusing on i-SDQ, we observe no substantial differences between results reported in Table \ref{tab_iSDQ} and those reported in Table \ref{tab_iSDQ_complete}.  Similarly, when focusing on e-SDQ, differences between estimates reported in Table \ref{tab_eSDQ} and \ref{tab_eSDQ_complete} may be considered as negligible. While further sensitivity analysis could be performed, based on these results, the MAR assumption seems to be quite reasonable. 

\section{Conclusions}\label{sec:6}
{The paper proposes a M-quantile regression for modelling location parameters of multivariate, continuous, longitudinal data. In particular, we extended the finite mixture of M-quantile regression models by \cite{AlfoRanalliSalvati2016} to account for multivariate longitudinal responses. Discrete, individual-specific, random effects were used to account for both dependence within the same response and association between responses, observed on the same sample unit.}
The model was applied to data on internalizing and externalizing SDQ  scores from the Millennium Cohort Study (MCS). In this application, we also handled the potential endogeneity of observed covariates by considering an auxiliary regression, in the spirit of \cite{AbrevayaDahl2008}. 
The results from the analysis of the MCS data are in line with those presented by \cite{Tzavidis2016}, but with some further insights that allow us to better characterize children's behavioural and emotional problems in terms of socio-economic conditions.

The proposed approach may be extended in a number of directions. First, we may consider time-varying random parameters in a hidden Markov model perspective, to capture possibly time-varying sources of unobserved heterogeneity in the data. Also, a further step may be taken by separately modelling the dependence between and within individual outcomes, with the aim at enhancing model flexibility. Last, an extension of the proposed approach to handle non-continuous responses represents a further direction to be investigated. 

\section*{Acknowledgments}
The work of Marino, Ranalli and Salvati has been developed under the support of the project PRIN-SURWEY (grant 2012F42NS8, Italy). The work of Alf\'{o} has been supported by the project ``Mixture and latent variable models for causal inference and analysis of socio-economic data'' (grant RBFR12SHVV, FIRB - Futuro in Ricerca).

%



\end{document}